# Healthcare Utilization, Chronic Condition Management, and Workplace Functioning Among Users of a Purpose-Built Mental Health AI (Ash): Cross-Sectional Study

Kristen M. Van Swearingen[1,2], Thomas D. Hull[2], Jeffrey Swigert[3], Caitlin A. Stamatis[2]

[1] Health Psychology Doctoral Program, The University of North Carolina at Charlotte, Charlotte, NC

[2] Slingshot AI, New York, NY

[3]University of Southern California's Center for Economic and Social Research (CESR)

**Corresponding Author:** Kristen M. Van Swearingen

**ORCID:** 0000-0002-8884-0770

**Abstract**

Mental health challenges can exacerbate physical symptoms and complicate management of chronic conditions. Purpose-built artificial intelligence (AI) tools may offer scalable support for co-occurring mental and physical health concerns. This cross-sectional study compared past-6-month healthcare utilization, chronic condition management, physical health behaviors, mental health change, and workplace functioning between active ($n = 169$) and non-users ($n = 73$) of a mental health AI (Ash). Participants had at least one chronic condition (e.g. hypertension, chronic pain). Binary outcomes were modeled as adjusted risk differences (RDs) using linear probability models and continuous outcomes were modeled with linear regression; all models were adjusted for hypertension. Relative to non-users, active users were more likely to report improved mental health (61.4% vs. 34.3%; RD = 0.27), higher medication adherence (91.7% vs. 76.4%, RD = 0.15), fewer skipped or delayed chronic-condition care activities ($b = -0.44$), and were less likely to report repeat urgent care visits (9.5% vs. 23.3%; RD = -0.15) and monthly-or-more absenteeism (24.2% vs. 45.2%; RD = -0.20, all $ps < .05$). Findings provide preliminary evidence that use of purpose-built AI may be associated with positive symptom-based and utilization outcomes for those managing co-occurring mental and physical concerns.

## Introduction

Mental health challenges can lead to, and be exacerbated by, physical health symptoms[1]. This relationship can be explained by both lifestyle and biological pathways. People with mental health challenges are more likely to exhibit unhealthy lifestyle behaviors, such as reduced physical activity, poor dietary patterns, and lower medication adherence, potentially mediated in part by a dysregulated stress response and heightened inflammation[2]. Relatedly, people with physical health conditions, such as cardiovascular disease, chronic pain, and diabetes, experience mental health concerns at higher rates than the general population[3–5].

The impact of comorbid mental and physical health concerns extends beyond symptom burden. Comorbid conditions are associated with worse disease management, higher healthcare utilization, and increased healthcare costs relative to either concern alone[1]. For instance, it is estimated that medical costs for those with comorbid physical and mental health conditions are two-to-three times that of those without a co-occurring mental health condition, amounting to $406 billion in excess annual healthcare spend in the United States[6]. This comorbidity also represents a known driver of costs to employers; as one example, working adults with arthritis and mental health challenges showed a 124% increased rate of absenteeism[7], and those with depression and cardiovascular disease were 8 times more likely than a healthy comparator group to report work impairment[8]. Given the impact of physical and mental health comorbidity on both individual outcomes and cost of care, there is an urgent need to enhance behavioral health support among those with chronic health conditions[9].

In light of challenges in scaling traditional human services (e.g., provider shortages, wait times, and financial costs[10–12], digital mental health tools have emerged as a potential means of augmenting behavioral support for people with chronic conditions. In principle, these tools are highly scalable and can be delivered at a fraction of the cost of human services. While one independent evaluation of digital mental health tools found that annual healthcare savings from improved depression and anxiety symptoms range from about $575 to $994 per user per year[13], an evaluation of digital diabetes solutions found a net *increase* in healthcare spend[14]. Together, these mixed findings leave open whether cost savings from digital health tools are more attributable to the health domain targeted, or to some other feature of intervention design. Moreover, whether tools built specifically for mental health could yield downstream physical health benefits among users with chronic conditions remains untested.

In recent years, there has been a proliferation of generative artificial intelligence (AI) tools used for mental health[15,16]. These tools vary considerably in design and oversight. Frontier models, built for general-purpose use, have documented safety limitations and can respond inadequately to users, particularly those in crisis situations or at risk for delusional thinking[17–19]. In contrast, purpose-built models are specifically designed for mental health use cases, incorporating clinical appropriateness standards and safety guardrails[20,21]. Purpose-built models have shown preliminary efficacy for depression and anxiety symptoms[20–22]. However, no study has examined whether AI tools demonstrated to improve depression and anxiety symptoms are also associated with fewer physical symptoms or cost savings among people with chronic medical conditions.

In the present study, we examined differences in physical and mental health outcomes among people with chronic medical conditions, comparing active users of a purpose-built AI for mental health (Ash) to non-users. We assessed group differences across five domains: (1) healthcare utilization, (2) physical health and health behaviors, (3) chronic condition management, (4) mental health, and (5) workplace functioning; we also estimated cost savings associated with active use. We hypothesized that relative to non-users, active users would report more favorable outcomes across each domain, including reduced acute care utilization, better adherence to chronic condition management, and improved workplace functioning. Given the exploratory design and the modest size of the non-user comparison group, we treated this study as an initial effort to estimate the magnitude of these associations to inform future, adequately powered research.

## Methods

**Participants and Procedures**

This was a cross-sectional study of real-world users of Ash (talktoash.com), a purpose-built conversational AI tool for mental well-being support. Participants had registered for Ash 6 months prior to survey administration and consented to data sharing for research purposes during onboarding. Two groups were targeted for recruitment based on mixpanel engagement data: (1) active users, who had engaged regularly with the chatbot (6 sessions or more) over the past 6 months, and (2) non-users, with 1 or fewer sessions over the same period. Eligible participants were invited via email to complete a questionnaire via Tally (tally.io) assessing healthcare utilization, physical health and health management, mental health, and workplace functioning. Data collection occurred in March 2026.

Among active registered users of Ash, 5,879 were emailed invitations to participate in the study, of which 461 (7.8%) opened the survey link, and 267 (4.5% of those initially emailed) completed the survey. Among non-users, 15,270 were emailed invitations to participate (a higher volume to account for an anticipated lower response rate), of which 265 (1.7%) opened the survey link, and 104 (0.7% of those initially emailed) completed the survey. In the present study, our analytic sample consisted of the 65.2% (242 out of 371 total) of participants who reported at least one diagnosed chronic health condition (e.g. type 2 diabetes or pre-diabetes, overweight or obesity, hypertension, hyperlipidemia, heart disease, chronic pain, asthma or chronic obstructive pulmonary disease, an autoimmune condition, or kidney disease), leading to a final analytic sample of 169 active users, and 73 non-users.

The study was reviewed by the Biomedical Research Alliance of New York (BRANY) independent Institutional Review Board (IRB) and determined to be exempt under category 4(ii) (BRANY IRB File # 26-081-2393). Survey respondents were entered into a drawing for one $500 gift card. As the initial response rate was low for non-users, in reminder emails, the non-user group was additionally offered a $10 gift card incentive for completing the survey.

**Measures**

***Demographics***

Participants provided information on their age (six-category bracket ranging from 18-24 to 65+), gender identity (categorized as man, woman, or non-binary/another gender), health insurance type (categorized as employer-sponsored, uninsured, Medicaid/Medicare/ACA Marketplace, or another type/undisclosed type), and lifetime mental health diagnosis (yes or no).

***Healthcare Utilization Measures***

Past-6-month visit frequency to a primary care provider, medical specialist, and psychiatrist were separately assessed on a 5-point ordinal scale (0 = Never, 4 = More than once a month [7+ times]). Past-6-month visit frequency to a therapist or counselor was assessed on a 6-point ordinal scale (0 = Never, 5 = More than once a week [25+ times total]). Past-6-month emergency department frequency was assessed on an ordinal scale (0 = never, 1 = once, 2 = twice, 3 = 3 or more times). These ordinal variables reflecting routine healthcare and emergency department visits were each coded as (0 = no visit, 1 = one or more visits) in the main analyses. Urgent care visit frequency in the past 6 months was assessed (0 = never, 1 = once, 2 = twice, 3 = 3 or more times), and coded as (0 = fewer than two visits, 1 = two or more visits) in the main analyses, reflecting repeat urgent care visits, as repeat visits may indicate an inadequately managed condition opposed to an isolated event. All were self-reported.

***Health Behaviors and Physical Health Measures***

Five single-item measures assessed health behaviors and physical health over the past 6 months. Participants rated sleep quality on a 5-point ordinal scale (1 = very poor, 5 = very good), which was coded as poor sleep quality (0 = "fair" or worse) or good sleep quality (1 = "good" or better). For exercise frequency, participants reported how many days per week they typically exercised for 30 or more minutes (0-7). Among those taking prescription medication for their chronic condition, medication adherence was reported on a 4-point ordinal scale (1 = rarely, 4 = always), which was coded as low medication adherence (0 = "sometimes" or less) or high medication adherence (1 = "most of the time" or more). Skipped or delayed care due to mental or emotional health was measured as a count (0-5) of which chronic-condition care activities (filling a prescription, attending a scheduled doctor's appointment, completing lab work or screening tests, exercising, and following a recommended diet) they had skipped or delayed. Physical health improvement was assessed with the item "Compared to 6 months ago, how would you describe your physical health?" rated on a 5-point ordinal scale (1 = much worse, 5 = much better), which was coded as not improved (0 = “about the same” or worse) or improved (1 = “somewhat better” or more).

***Mental Health Measures***

Depression and anxiety were assessed using the 2-item Patient Health Questionnaire (PHQ-2)[23] and 2-item Generalized Anxiety Disorder scale (GAD-2)[24], respectively; participants reported how often they were bothered by each symptom over the past 2 weeks; scores ranged from 0-6 with higher scores indicating greater symptoms. Mental health improvement was assessed with the item "Compared to 6 months ago, how would you describe your mental/emotional health?" rated on a 5-point ordinal scale (1 = much worse, 5 = much better), which was coded as not improved (0 = “about the same” or worse) or improved (1 = “somewhat better” or more). Participants also reported the extent to which their daily activities were affected by mental or emotional health over the past 6 months, on a 5-point ordinal scale (0 = not at all, 4 = extremely), which was coded as low daily activity impact (0 = “a little” or less) or moderate+ daily activity impact (1 = “moderately” or more).

***Workplace and Daily Functioning Measures***

Presenteeism was assessed with the item “In the past 6 months, how often were you significantly less productive at work or school due to mental or emotional health” rated as (0 = never, 4 = very often [multiple days a week]), which was coded as low presenteeism (0 = “a few days total” or less) and monthly+ presenteeism (1 = “a few days a month” or more). Absenteeism was assessed with the item “In the past 6 months, how often did you miss days of work or school due to mental or emotional health?” reported as (0 = never, 4 = very often [multiple days a week]), which was coded as (0 = less than monthly, 1 = monthly or more).

**Statistical Analyses**

All analyses were conducted in R version 4.4.1[25] with base R for main analyses, *sandwich*[26] for heteroskedasticity-robust (HC3) covariance estimation and *lmtest*[27] for coefficient tests and confidence intervals. We report descriptive statistics for the full analytic sample and separately by active and non-user subgroups. Group differences in demographic characteristics (age, gender, insurance type) and chronic condition prevalence were tested using chi-square tests of independence for categorical variables with more than two levels and two-proportion z-tests for binary characteristics, including lifetime mental health diagnosis and each individually reported chronic condition. Because active and non-users differed in their prevalence of hypertension, we therefore added hypertension as a covariate to our primary analyses. Binary outcomes were modeled as adjusted risk differences (RD) using linear probability models with heteroskedasticity-robust (HC3) standard errors. Outcomes that were count variables and symptom-score metrics (chronic-care activities that were skipped or delayed, PHQ-2, GAD-2, and weekly exercise days) were modeled as linear regression and reported as adjusted mean differences. As a sensitivity analysis, ordinal outcomes that were dichotomized in the primary models were assessed with their full response scales using ordinal logistic regression to assess whether conclusions drawn from the primary analyses were dependent on the pre-specified dichotomization. Categories were collapsed if necessary to retain at least 10 responses per category.

## Results

### Study Participants and Descriptive Data

The study sample included 242 participants with at least one chronic health condition, including 169 (69.8%) active users, and 73 (30.2%) non-users. Table 1 presents demographic characteristics for the entire analytic sample and by engagement group. Participants were predominantly female (73.1%), age 35-44 (29.8%), had health insurance through Medicaid, Medicare, or the ACA Marketplace (44.6%), and had a lifetime diagnosis of a mental health condition (68.6%). Active and non-users did not differ significantly in age, gender, health insurance type, or lifetime mental health diagnosis (*ps* > .05). For chronic conditions, 57.4% reported overweight or obesity (BMI ≥ 25), 40.9% reported chronic pain, 33.9% reported high blood pressure, and 24.0% reported type 2 diabetes or pre-diabetes. Active and non-users differed significantly in the prevalence of high blood pressure/hypertension ($z = -2.45$, $p = .014$); no other chronic condition differed significantly between groups (*ps* > .24). See Table 2 for the full list of reported chronic conditions. In the overall sample, 38.0% ($n$ = 92) reported just one chronic condition, 25.2% ($n$ = 61) reported two chronic conditions, and 36.8% ($n$ = 89) reported three or more chronic conditions.

**Table 1**

*Demographic Characteristics by Engagement Group*

| Characteristic | Full Sample (N = 242) | Active Users (n = 169) | Non-Users (n = 73) |
|---|---|---|---|
| **Age** | | | |
| 18-24 | 15 (6.2%) | 9 (5.3%) | 6 (8.2%) |
| 25-34 | 43 (17.8%) | 34 (20.1%) | 9 (12.3%) |
| 35-44 | 72 (29.8%) | 41 (24.3%) | 31 (42.5%) |
| 45-54 | 43 (17.8%) | 34 (20.1%) | 9 (12.3%) |
| 55-64 | 39 (16.1%) | 30 (17.8%) | 9 (12.3%) |
| 65+ | 16 (6.6%) | 10 (5.9%) | 6 (8.2%) |
| No response | 14 (5.8%) | 11 (6.5%) | 3 (4.1%) |
| **Gender** | | | |
| Man | 44 (18.2%) | 27 (16.0%) | 17 (23.3%) |
| Woman | 177 (73.1%) | 126 (74.6%) | 51 (69.9%) |
| Non-binary/another gender | 7 (2.9%) | 5 (3.0%) | 2 (2.7%) |
| No response | 14 (5.8%) | 11 (6.5%) | 3 (4.1%) |
| **Health Insurance Type** | | | |
| Employer-sponsored | 79 (32.6%) | 60 (35.5%) | 19 (26.0%) |
| Uninsured | 18 (7.4%) | 13 (7.7%) | 5 (6.8%) |
| Medicaid/Medicare/ACA Marketplace | 108 (44.6%) | 67 (39.6%) | 41 (56.2%) |
| Other/prefer not to say | 23 (9.5%) | 18 (10.7%) | 5 (6.8%) |
| No response | 14 (5.8%) | 11 (6.5%) | 3 (4.1%) |
| **Lifetime Mental Health Diagnosis** | | | |
| Yes | 166 (68.6%) | 115 (68.0%) | 51 (69.9%) |
| No | 52 (21.5%) | 37 (21.9%) | 15 (20.5%) |

| No response | 24 (9.9%) | 17 (10.1%) | 7 (9.6%) |
|---|---|---|---|

Note. Values are n (%). Percentages are based on the full analytic sample (N = 242; active users n = 169, non-users n = 73). Participants who did not answer an item are shown as "No response." Active users and non-users did not differ significantly in age, gender, insurance type, or lifetime mental health diagnosis (all *ps* > .05).

**Table 2**

*Chronic Health Conditions Reported by Engagement Group*

| **Chronic Condition** | **Full Sample (N = 242)** | **Active Users (n = 169)** | **Non-Users (n = 73)** |
|---|---|---|---|
| Obesity/overweight (BMI ≥ 25) | 139 (57.4%) | 101 (59.8%) | 38 (52.1%) |
| High blood pressure/hypertension[a] | 82 (33.9%) | 49 (29.0%) | 33 (45.2%) |
| Chronic pain | 99 (40.9%) | 68 (40.2%) | 31 (42.5%) |
| Type 2 diabetes or pre-diabetes | 58 (24.0%) | 42 (24.9%) | 16 (21.9%) |
| High cholesterol/hyperlipidemia | 57 (23.6%) | 41 (24.3%) | 16 (21.9%) |
| Asthma/COPD | 56 (23.1%) | 40 (23.7%) | 16 (21.9%) |
| Autoimmune condition | 40 (16.5%) | 29 (17.2%) | 11 (15.1%) |
| Heart disease | 21 (8.7%) | 17 (10.1%) | 4 (5.5%) |
| Kidney disease | 9 (3.7%) | 7 (4.1%) | 2 (2.7%) |

Note. Conditions were assessed with a multi-select item ("Have you been diagnosed with any of the following?"); respondents could select more than one condition (*M* = 2.32 conditions per person, range = 1-7), so percentages do not sum to 100%.

aActive and non-users differed significantly in the prevalence of high blood pressure/hypertension ($z = -2.45$, $p = .014$); no other condition differed significantly between groups (all *ps* > .24).

**Outcome Comparisons by Domain**

Table 3 presents comparisons of healthcare utilization, health behaviors and physical health, mental health, and workplace functioning outcomes between active and non-users within the analytic sample of participants with chronic health conditions. Analyses were adjusted for hypertension status.

### *Healthcare Utilization*

Active and non-users did not differ significantly in the likelihood of any past-6-month visit to a primary care provider, medical specialist, therapist or counselor, or psychiatrist (*ps* > .44). Active users were significantly less likely to report repeat (two or more) urgent care visits (9.5% vs. 23.3%; RD = -0.15, $p$ = 007). The likelihood of any emergency department visit did not differ significantly between groups ($p$ = .702)

### *Health Behaviors and Physical Health*

Active users were significantly more likely than non-users to report improvement in their physical health over the past 6-months (36.9% vs. 23.3%, RD = 0.14, $p$ = .027), and to report high medication adherence (i.e. taking medications “most of the time” or “always”) (91.7% vs. 76.4%; RD = 0.15, $p$ = .018). The likelihood of good sleep quality did not differ significantly between groups ($p$ = .321), nor did weekly exercise frequency ($p$ = .657). Active users reported that they skipped or delayed significantly fewer chronic-condition care activities due to mental or emotional health ($M$ = 1.11, $SD$ = 1.25) than non-users ($M$ = 1.55, $SD$ = 1.49; $b$ = -0.44, $p$ = .033), and were significantly more likely to report that their mental or emotional health did not affect any chronic-condition care activity (45.0% vs. 28.8%; RD = 0.17, $p$ = .012). This group difference was primarily driven by medication adherence, with non-users significantly more likely to report skipping or delaying a prescription refill (26.0% vs. 12.4%; RD = -0.14, $p$ = .019). Table 4 presents mental health impacts on individual chronic-care activities by engagement group.

### *Mental Health*

Active users reported significantly lower depressive symptoms over the past two weeks on the PHQ-2 ($M$ = 2.01, $SD$ = 1.75) than non-users ($M$ = 2.76, $SD$ = 1.98; $b$ = -0.73, $p$ = .009). GAD-2 scores did not differ significantly between groups ($p$ = .114). Active users were significantly more likely than non-users to report improvement in their mental health over the past 6-months (61.4% vs. 34.3%; RD = 0.27, $p < .001$). The likelihood that mental or emotional health moderately or severely affected daily activities did not differ significantly between groups ($p$ = .337).

### *Workplace Functioning*

Active users were significantly less likely than non-users to report monthly-or-more absenteeism due to mental or emotional health (24.2% vs. 45.2%; RD = -0.20, $p$ = .004). Monthly-or-more presenteeism trended lower among active users (54.3% vs. 65.8%; RD = -0.13, $p$ = .055).

**Table 3**

*Outcome Comparisons by Engagement Group, Adjusted for Hypertension*

| Outcome | Active User (n = 169) | Non-User (n = 73) | n (active / non-user) | Adjusted RD / b | 95% CI | *p* |
|---|---|---|---|---|---|---|
| ***Healthcare Utilization*** | | | | | | |
| Any primary care visit, n (%) | 134 (79.3) | 60 (82.2) | 169 / 73 | RD = -0.01 | [-0.12, 0.10] | 0.805 |
| Any specialist visit, n (%) | 88 (52.1) | 42 (57.5) | 169 / 73 | RD = -0.05 | [-0.19, 0.09] | 0.484 |
| Any therapist visit, n (%) | 90 (53.3) | 38 (52.1) | 169 / 73 | RD = 0.02 | [-0.12, 0.16] | 0.810 |
| Any psychiatrist visit, n (%) | 64 (37.9) | 32 (43.8) | 169 / 73 | RD = -0.05 | [-0.19, 0.09] | 0.449 |
| Any emergency department visit, n (%) | 58 (34.3) | 27 (37.0) | 169 / 73 | RD = -0.03 | [-0.16, 0.11] | 0.702 |
| Repeat urgent care (2+), n (%) | 16 (9.5) | 17 (23.3) | 169 / 73 | RD = -0.15 | [-0.26, -0.04] | 0.007 |
| ***Health Behaviors and Physical Health*** | | | | | | |
| Physical health improved, n (%) | 62 (36.9) | 17 (23.3) | 168 / 73 | RD = 0.14 | [0.02, 0.27] | 0.027 |
| Care activities skipped/delayed, M (SD) | 1.11 (1.25) | 1.55 (1.49) | 169 / 73 | b = -0.44 | [-0.84, -0.04] | 0.033 |
| High medication adherence, n (%) [a] | 122 (91.7) | 42 (76.4) | 133 / 55 | RD = 0.15 | [0.03, 0.27] | 0.018 |
| Good sleep quality, n (%) | 46 (27.4) | 14 (19.2) | 168 / 73 | RD = 0.06 | [-0.06, 0.17] | 0.321 |
| Exercise days/week, M (SD) | 2.11 (2.08) | 2.18 (2.29) | 169 / 73 | b = -0.14 | [-0.76, 0.48] | 0.657 |
| ***Mental Health*** | | | | | | |
| Mental health improved, n (%) | 97 (61.4) | 24 (34.3) | 158 / 70 | RD = 0.27 | [0.13, 0.41] | < .001 |
| PHQ-2 total, M (SD) | 2.01 (1.75) | 2.76 (1.98) | 160 / 71 | b = -0.73 | [-1.29, −0.18] | 0.009 |
| GAD-2 total, M (SD) | 2.59 (1.87) | 3.01 (1.85) | 157 / 70 | b = -0.43 | [-0.97, 0.11] | 0.114 |
| Moderate+ daily activities impact, n (%) | 82 (50.0) | 42 (57.5) | 164 / 73 | RD = -0.07 | [-0.21, 0.07] | 0.337 |
| ***Workplace Functioning*** | | | | | | |
| Monthly+ absenteeism, n (%) | 40 (24.2) | 33 (45.2) | 165 / 73 | RD = -0.20 | [-0.34, −0.07] | 0.004 |
| Monthly+ presenteeism, n (%) | 89 (54.3) | 48 (65.8) | 164 / 73 | RD = -0.13 | [-0.27, 0.00] | 0.055 |

Note. Binary outcomes are reported as adjusted risk differences (RD) from linear probability models; continuous outcomes (care activities skipped/delayed, PHQ-2, GAD-2, exercise days) are reported as adjusted mean differences (b) from linear regression. All models included active-user status as the predictor and hypertension (1 = diagnosed hypertension, 0 = no hypertension) as a covariate, with heteroskedasticity-robust (HC3) standard errors. RD is the adjusted difference in the probability of the outcome with positive values indicating a higher probability among active users. Descriptive n (%) and M (SD) are unadjusted.

[a] Medication adherence was assessed only among respondents prescribed chronic-condition medications.

**Table 4**

*Health-Related Activities That Were Skipped or Delayed Due to Mental or Emotional Health by Engagement Group*

| Skipped or Delayed Activity | Full Sample (N = 242) | Active Users (n = 169) | Non-Users (n = 73) | RD [95% CI] | *p* |
|---|---|---|---|---|---|
| Exercise or physical activity routine | 111 (45.9%) | 72 (42.6%) | 39 (53.4%) | -0.12 [-0.26, 0.02] | 0.096 |
| A scheduled doctor's appointment for a chronic condition | 58 (24.0%) | 35 (20.7%) | 23 (31.5%) | -0.10 [-0.23, 0.02] | 0.104 |
| Following a recommended diet | 53 (21.9%) | 33 (19.5%) | 20 (27.4%) | -0.08 [-0.20, 0.04] | 0.196 |
| Lab work or screening tests | 43 (17.8%) | 28 (16.6%) | 15 (20.5%) | -0.04 [-0.14, 0.07] | 0.512 |
| Filling a prescription | 40 (16.5%) | 21 (12.4%) | 19 (26.0%) | -0.14 [-0.25, -0.02] | 0.019 |
| None- mental health did not affect these | 97 (40.1%) | 76 (45.0%) | 21 (28.8%) | 0.17 [0.04, 0.30] | 0.012 |

Note. Assessed with a multi-select item ("In the past 6 months, did you skip or delay any of the following due to mental or emotional health?"); respondents could select more than one activity. Values are adjusted risk differences (RD) with 95% CIs from linear probability models comparing active versus non-users, with active-user status as the predictor and hypertension (1 = diagnosed hypertension, 0 = no hypertension) as a covariate; a negative RD indicates a lower probability among active users.

### Sensitivity Analyses

Ordinal logistic regression sensitivity analyses on the full response scale for variables that were dichotomized were largely consistent with the primary findings (Supplementary Table S1). Routine care variables remained non-significant, absenteeism and medication adherence remained significant, and urgent care attenuated to a strong trend.

## Discussion

The present study compared active and non-users of a purpose-built conversational AI tool for mental health (Ash) among adults with chronic health conditions. Compared to non-users, active users were more likely to report improvement in both mental and physical health over the past 6 months, alongside less frequent acute care utilization and more consistent chronic condition management behaviors. Active users also reported lower absenteeism from work or school due to mental or emotional health. Taken together, these findings suggest that active engagement with a purpose-built AI for mental health support may be associated with broad downstream outcomes spanning behavioral health, physical health, care utilization and management, and workplace functioning.

The study extends a growing body of work on purpose-built conversational AI for mental health, which has largely focused on mental health outcomes among general populations[20,28,29]. One prior purpose-built AI for mental health study examined clinical outcomes within a sample of users managing chronic pain[22]. However, to our knowledge, this is the first study to examine healthcare utilization and chronic condition management among users of a purpose-built mental health AI tool who were also managing a chronic health condition. This is a particularly relevant population to study, given that mental health concerns frequently impact physical health, and comorbid mental and physical health conditions are associated with worse disease management and higher healthcare costs[1,6]. Participants in the current study reported high rates of chronic health conditions (65.2%), broadly consistent with national data indicating that 59.5% of adults aged 18-34 and 78.4% of adults aged 35-64 report at least one chronic health condition[30]. This suggests that the population studied here is not a narrow or unusual one, though whether these specific findings generalize to the broader population of adults with chronic conditions—most of whom have not used a mental health app like Ash—remains to be tested directly.

Purpose-built AI tools for mental health have shown preliminary effectiveness for depression and anxiety[20,21,29], and the current findings align with this work. Depression symptoms were 27% lower among active users compared to non-users, and active users were nearly twice as likely to report improved mental health over the past 6 months (61.4% vs. 34.3%). Active and non-users did not differ detectably in their routine or mental health care over the past 6 months; although the data is not able to establish equivalence between groups, this pattern suggests that active users were not simply more engaged with the healthcare system overall.

Although active and non-users did not differ detectably in routine healthcare use, they differed in acute healthcare utilization: non-users were more than twice as likely to report repeat (two or more) urgent care visits in the past 6 months (9.5% vs. 23.3%). Given a 1,000 employee workforce, this corresponds with an estimated $56,572 difference in annual costs (see Supplementary Material Table S2). Urgent care centers often bridge the gap between routine and emergency care for people managing chronic conditions[31], and repeat visits may signal inadequately managed chronic illness rather than isolated acute events. While the sensitivity analysis using the full ordinal scale (Supplementary Material Table S1) was directionally consistent with the primary analysis, it did not reach significance ($p = .053$), which is expected when group differences are not distributed evenly and are more heavily concentrated in the sparse categories. As there is a meaningful distinction in repeat urgent care visits over a single instance, the primary analysis showing differences in repeat care is still a meaningful finding.

Active users were more likely to report improvement in physical health over the past 6 months than non-users (36.9% vs. 23.3%), a 58% relative difference. Two plausible pathways may explain this pattern of better physical health. The first is physiological, where improved mental health can help to regulate the stress response and reduce inflammation[32,33] and dysregulation of these systems which can drive the onset and exacerbation of physical symptoms[32,34]. Because the present study did not measure physiological indicators, this pathway remains speculative. The second pathway is behavioral, and is supported by our findings on chronic condition management. Active users reported skipping or delaying significantly fewer chronic-condition care activities due to mental health than non-users, who were more than twice as likely to delay filling a prescription. Active users were also more likely to report high medication adherence. Treatment non-adherence is a well-established driver of symptom exacerbation, complications, acute care utilization, and avoidable healthcare spend[35,36], particularly for the comorbid conditions represented in our sample. Mental health challenges are recognized as a major contributor to poor adherence[37,38]. Schönenberg and collaborators[38], for instance, found that depression symptoms, particularly problems with concentration, decision making, and fatigue, were significant predictors of non-adherence even controlling for sociodemographic and health covariates. This suggests that improved mental health may support physical health not only through physiological pathways, but also by strengthening the behavioral capacity to manage a chronic condition. Notably, weekly exercise frequency and sleep quality did not differ significantly between groups suggesting that the observed benefits may be more closely tied to chronic condition management behaviors rather than broader lifestyle behavior change. Taken together, lower acute care use alongside greater perceived physical health improvement, without a corresponding increase in routine care, is consistent with the possibility that mental health support may offer measurable downstream benefits for physical health, though the cross-sectional design cannot establish causality.

Downstream effects of mental health support may extend beyond health outcomes to how people function day to day. Both chronic conditions and mental health concerns are associated with poorer workplace functioning[7,8], and prior studies of digital mental health tools have shown improvements in workplace outcomes, such as productivity and work impairment, following behavioral health support[39,40], though evidence across this literature is mixed [41]. In the current study, active users reported lower absenteeism due to mental or emotional health than non-users. With the full scale measure, the difference in absenteeism corresponds to a difference of about 3.60 estimated missed days over 6 months. This could translate to an estimated $1,017,798 in annual savings for a hypothetical 1,000-employee workforce (see

Supplementary Material Table S2). Presenteeism, however, only marginally differed between groups. This dissociation may reflect what each measure captures: absenteeism indexes the capacity to show up at all[42], which is plausibly responsive to improved mood and stress regulation, whereas at-work productivity may be more heavily constrained by job demands, workload, and organizational factors that mental health support alone may not change[43]. This interpretation remains speculative, and workplace policies and job expectations may have contributed to both findings.

These findings should be weighed against both the design features that strengthen them and the constraints that limit their interpretation. The study extends prior work on purpose-built conversational AI to healthcare utilization and chronic condition management, and captures real-world engagement with strong ecological validity. The study also draws on a comparison group similar to active users across demographics, insurance type, mental health diagnosis, and observed routine care utilization, which argues against differential healthcare engagement as an explanation for group differences, though we can not rule it out. At the same time, this was a cross-sectional study of self-selected groups, precluding causal conclusions and leaving unmeasured possible differences such as baseline severity and health motivation. As these groups were distinguished based on their level of engagement, findings may be vulnerable to the "healthy adherer" effect, in which active engagement may signal better health-promoting behaviors, potentially biasing the associations observed in the study[44]. Completion rates were low, particularly among non-users (0.7% vs 4.5% of those invited), and all participants had registered for Ash, so respondents may differ from both non-respondents and adults with chronic conditions generally. Additionally, measures were self-reported over a 6-month recall period, and active users may have been motivated to report improvement. Given the modest size of the non-user subgroup (n = 73) and limited power, we treated this study as an effort to estimate effect sizes to inform a future, adequately powered study, and interpreted findings in light of effect sizes and the pattern across outcomes rather than individual tests; and the cost estimates should be treated as a plausible range. Prospective designs and linkage to claims data would allow stronger causal inference and independent validation of self-reported utilization.

To our knowledge, this is the first study to examine healthcare utilization and chronic condition management among users of a purpose-built conversational AI for mental health. Among a real-world sample with naturalistic engagement, active users reported better outcomes spanning mental health, physical health, acute care utilization, chronic condition management, and workplace functioning relative to those who had registered for the app but remained non-active. These associations cannot be interpreted causally given the cross-sectional design and self-selected comparison groups. Longitudinal, claims-based research remains an important next step. Still, given the need to enhance behavioral health support for people with chronic conditions, the consistency of the differences observed here across largely independent outcome domains suggests that purpose-built, scalable tools such as Ash may offer meaningful support for working-age adults managing co-occurring mental and physical health concerns.

**Data Availability Statement**

Data are available by reasonable request and a data use agreement.

**Author Contributions Statement**

K.V.S. performed the data analysis and drafted the manuscript. T.D.H. supervised the research. J.S. provided data analysis guidance. C.A.S. designed the study, collected the data, performed the data analysis, drafted the manuscript, and supervised the research. All authors reviewed and edited the manuscript.

**Declaration of Competing Interests**

CAS and TDH are employees of the company building the model and declare no non-financial competing interests. KVS received a stipend from the company to contribute to this research and declares no other competing interests. JS declares no competing interests.

## Supplemental Materials

**Table S1**

*Sensitivity Analysis: Ordinal Logistic Regression on Full Response Scales, Adjusted for Hypertension*

| Outcome | Proportional OR | 95% CI | *p* | n |
|---|---|---|---|---|
| Primary care visits | 1.36 | [0.82, 2.24] | 0.229 | 242 |
| Specialist visits | 0.93 | [0.56, 1.55] | 0.791 | 242 |
| Therapist/counselor visits | 1.11 | [0.67, 1.86] | 0.684 | 242 |
| Psychiatrist visits | 0.77 | [0.45, 1.32] | 0.342 | 242 |
| Emergency department visits | 0.84 | [0.48, 1.48] | 0.548 | 242 |
| Urgent care visits | 0.58 | [0.33, 1.01] | 0.053 | 242 |
| Medication adherence | 1.90 | [1.02, 3.56] | 0.043 | 188 |
| Sleep quality | 1.60 | [0.96, 2.65] | 0.070 | 241 |
| Daily activities impact | 0.65 | [0.39, 1.09] | 0.103 | 237 |
| Absenteeism | 0.45 | [0.27, 0.76] | 0.003 | 238 |
| Presenteeism | 0.66 | [0.40, 1.09] | 0.106 | 237 |

Note. Proportional odds ratios (ORs) are from ordinal logistic regression models fit on each outcome's full ordinal response scale, with active-user status as the predictor and hypertension (1 = diagnosed hypertension, 0 = no hypertension) as a covariate. A proportional OR greater than 1 indicates higher odds of a more frequent or more favorable response category among active users relative to non-users while an OR less than 1 indicates lower odds. Sparse upper categories were collapsed to retain at least 10 observations each. Corresponding primary-analysis estimates (adjusted risk differences) appear in Table 3 of the main text.

### *Cost Analysis Estimation*

Because acute care visits and missed workdays both carry direct, quantifiable costs, these two findings allow a preliminary estimate of the economic implications of active engagement with Ash. Whereas the primary analyses modeled urgent care visits and absenteeism as binary outcomes, the cost estimates presented here used the full response scale of these outcomes in an ordinary least squares regression adjusted for hypertension, so that the adjusted differences between active and non-users could be expressed as the difference in the expected number of urgent care visits and missed workdays. All estimates use a hypothetical 1,000-employee workforce, assuming a 65.2% chronic-condition prevalence (as found in our sample, and consistent with the national average reported by the Centers for Disease Control [CDC][1]. Specifically, the CDC reports that 59.5% of adults aged 18-34, and 78.4% aged 35-64, report at least one chronic health condition.

Urgent care visits over the past 6 months were measured on the following scale (0 = none, 1 = once, 2 = twice, 3 = three or more times). Active users averaged fewer urgent care visits than non-users (0.50 visits vs. 0.74 visits). Lower urgent care utilization reported by active users corresponded to an adjusted difference of 0.26 fewer visits per person over 6 months (95% CI: 0.02 - 0.50). Assuming an average urgent care visit cost of $164[2], this translates to an estimated $56,572 (95% CI: $5,192 - $107,952) in annual savings. Estimated savings from reduced absenteeism were substantially larger. Absenteeism visits over the past 6 months were measured on the following scale ("never," "rarely [1–2 days total]," "sometimes [about once a month]," "often [2–3 times a month]," "very often [once a week or more]") which were mapped to an estimated number of missed days over 6-months (never = 0, rarely = 1.5, sometimes = 6, often = 15, very often = 26). In a sensitivity analysis restricted to adults aged 25 and older (n = 213), for whom missed days more plausibly reflect work than school, active users reported an adjusted 3.16 fewer missed workdays per person over 6 months (95% CI: 0.96 - 5.36; $p = .005$). Applying a $247 median daily wage[3], this corresponds to an estimated average $1,017,798 (95% CI: $307,965 - $1,727,631) in annual savings. Table 2 summarizes both estimates. These estimates rest on several modeling assumptions (e.g., national rather than employer-specific benchmarks; mapping of an ordinal absenteeism frequency scale to estimated day counts) and should be interpreted as a plausible range under stated assumptions. Because the study was cross-sectional, these estimates reflect associations rather than causal effects of Ash engagement. Concurrent measurement of engagement and outcomes, substantial differential nonresponse, and absence of baseline outcome data preclude causal interpretation.

**Table S2**

*Estimated Annual Cost Savings Associated with Active Engagement*

| Cost driver | Adjusted difference per person, 6 months (95% CI) | Estimated Unit cost | Estimated annual savings (95% CI) |
|---|---|---|---|
| Urgent care visits | 0.26 fewer visits (0.02 - 0.50) | $164 per visit | $56,572 ($5,192 - $107,952) |
| Work absenteeism (adults ≥ 25) | 3.16 fewer missed workdays (0.96 - 5.36) | $247 per day | $1,017,798 ($307,965 - $1,727,631) |

Note. Estimates derive from hypertension-adjusted regression models among respondents with a chronic condition, scaled to 652 affected employees (65.2% of 1,000).